\documentclass[twocolumn]{revtex4-1}

\usepackage{graphicx}
\usepackage{dcolumn}
\usepackage{amsmath}

\begin{document}

\title{Renormalization group analysis of the hyperbolic sine-Gordon model\\
-- Asymptotic freedom from cosh interaction -- 
}

\author{Takashi Yanagisawa
}

\affiliation{Electronics and Photonics Research Institute,
National Institute of Advanced Industrial Science and Technology (AIST),
Tsukuba Central 2, 1-1-1 Umezono, Tsukuba 305-8568, Japan
}


\begin{abstract}
We present a renormalization group analysis for the hyperbolic sine-Gordon (sinh-Gordon)
model in two dimensions.
We derive the renormalization group equations based on the dimensional
regularization method and the Wilson method.  The same equations are obtained 
using both these methods.
We have two parameters $\alpha$ and $\beta\equiv \sqrt{t}$ where $\alpha$ indicates
the strength of interaction of a real salar field and $t=\beta^2$ is
related with the normalization of the action. 
We show that $\alpha$ is renormalized to zero in the high-energy region, that is, 
the sinh-Gordon theory is an asymptotically free theory. 
We also show a non-renormalization property that the beta function of $t$
vanishes in two dimensions. 
\end{abstract}


\maketitle

\section{Introduction}

The sine-Gordon model is an important  model and plays a significant role
in physics\cite{col75,das75,jos77,sam78,zam79,ami80,wie78,bal00,raj82,man04,
col85,mar17,wei15,her07}.  
It is known that
the sine-Gordon model is equivalent to the massive Thirring model in the
weak coupling phase\cite{col75,man75,sch77,fab01}.
The sine-Gordon model has universality in the sense that there are
many phenomena that are closely related to it.
The two-dimensional sine-Gordon model is mapped to the Coulomb gas model
with logarithmic Coulomb interaction\cite{sam78,jos76,zin89}.

The hyperbolic sine-Gordon model (sinh-Gordon model) is similar to 
the sine-Gordon model, where the cosine potential is replaced by the
hyperbolic cosine one.
The sinh-Gordon model has been studied as a field theory
model\cite{ing86,kou93,fri93,waz05,byt06,hoe07,tes08}.
It appears that the sinh-Godron model is similar to the $\phi^4$
model when we expand $\cosh\phi$ in terms of $\phi$.
In fact, both models have a kink solution as a classical solution.
There is, however, the significant difference between these models that
the $\phi^4$ model is renormalized in four dimensions while the
sinh-Gordon model is renormalizable in two dimensions.

The Lagrangian of the sine-Gordon model is given as
\begin{equation}
\mathcal{L}= \frac{1}{2t}(\partial_{\mu}\phi)^2
+\frac{\alpha}{t}\cos\phi,
\end{equation}
for a real scalar field $\phi$.
This is written as
\begin{equation}
\mathcal{L}= \frac{1}{2}(\partial_{\mu}\phi)^2
+g\cos(\beta\phi),
\end{equation}
where $\beta=\sqrt{t}$ and $g=\alpha/t$ with the transformation
$\phi\rightarrow \beta\phi$. 
The sinh-Gordon model is obtained by performing the transformation
in Eq. (2):
\begin{eqnarray}
\beta &\rightarrow& i\beta, \\
g &\rightarrow& -g.
\end{eqnarray}

In this paper we investigate the sinh-Gordon model by using the
renormalization group theory.  We use the dimensional regularization
method\cite{tho72,gro76} as well as the Wilson renormalization group 
method\cite{wil75,kog79}.
The beta functions are derived using these methods and show that
the coupling constant for the hyperbolic cosine potential decreases
as the energy scale increases.  Namely, the model shows an
asymptotic freedom.

The paper is organized as follows.
In Sect. 2, we present the model that we consider in this paper.
In Sect. 3, we derive renormalization group equations on the
basis of the dimensional regularization method.
In Sect. 4, we examine the renormalization procedure based on the
Wilson method, and in Sect. 5 we investigate the scaling property.
In Sect. 6 we consider the generalized model with high-frequency
modes and examine their effect on scaling property.
A summary is given in the final section.

\section{Sinh-Gordon model}

We consider the Lagrangian density for a real scalar field $\phi$:
\begin{equation}
\mathcal{L}= \frac{1}{2t_0}(\partial_{\mu}\phi_B)^2
-\frac{\alpha_0}{t_0}\cosh(\phi_B),
\end{equation}
where $t_0$ and $\alpha_0$ are bare coupling constants, and $\phi_B$ is a
bare real scalar field. The second term is the potential energy given
by the hyperbolic cosine function $\cosh\phi=(e^{\phi}+e^{-\phi})/2$.
$t$ and $\alpha$ denote the renormalized coupling constants.  They are related
to bare quantities through the relations given as
\begin{eqnarray}
t_0&=& t\mu^{2-d}Z_t,\\
\alpha_0 &=& \alpha\mu^2 Z_{\alpha},
\end{eqnarray}
where $Z_t$ and $Z_{\alpha}$ are renormalization constants.
We introduced the energy scale $\mu$ so that $t$ and $\alpha$ are 
dimensionless constants.
We adopt that $t$ and $\alpha$ are positive: $t>0$ and $\alpha>0$.
The renormalized field $\phi_R$ is defined by
\begin{equation}
\phi_B= \sqrt{Z_{\phi}}\phi_R,
\end{equation}
where we introduced the renormalization constant $Z_{\phi}$ for the
field $\phi$.
Then, the Lagrangian with renormalized quantities is
\begin{equation}
\mathcal{L}= \frac{\mu^{d-2}Z_{\phi}}{2tZ_t}(\partial_{\mu}\phi)^2
-\frac{\mu^d\alpha Z_{\alpha}}{tZ_t}\cosh(\sqrt{Z_{\phi}}\phi),
\end{equation}
where $\phi$ indicates the renormalized field $\phi_R$.

We consider the Euclidean action for convenience.  The action for
the sinh-Gordon model in $d$ dimensions reads
\begin{equation}
S= \int d^dx\Big[ \frac{\mu^{d-2}Z_{\phi}}{2tZ_t}(\partial_{\mu}\phi)^2
+\frac{\mu^d\alpha Z_{\alpha}}{tZ_t}\cosh(\sqrt{Z_{\phi}}\phi)\Big].
\end{equation}

\section{Renormalization group equations}

\subsection{Renormalization of $\alpha$}

We consider tadpole diagrams to take account of the renormalization
of $\alpha$ up to the lowest order of $\alpha$ (Fig. 1)\cite{ami80,yan16,yan17,yan18}.
Using the expansion $\cosh\phi=1+(1/2)\phi^2+(1/4!)\phi^4+\cdots$,
the hyperbolic cosine function is renormalized  in a similar way
to the sine-Gordon model, as
\begin{eqnarray}
\cosh(\sqrt{Z_{\phi}}\phi) &\rightarrow& \left( 1+\frac{1}{2}Z_{\phi}
\langle\phi^2\rangle+\cdots \right)\cosh(\sqrt{Z_{\phi}}\phi)
\nonumber\\
&=& \exp\left( \frac{1}{2}Z_{\phi}\langle\phi^2\rangle\right)
\cosh(\sqrt{Z_{\phi}}\phi).
\end{eqnarray}
The expectation value $\langle\phi^2\rangle$ is evaluated as
\begin{eqnarray}
Z_{\phi}\langle\phi^2\rangle &=& t\mu^{2-d}Z_t\int\frac{d^dk}{(2\pi)^d}
\frac{1}{k^2+m_0^2} \nonumber\\
&=& -\frac{1}{\epsilon}t\frac{\Omega_d}{(2\pi)^d},
\end{eqnarray}
where we put
\begin{equation}
d=2+\epsilon.
\end{equation}
$m_0$ was introduced to avoid the infrared divergence and $\Omega_d$
is the solid angle in $d$ dimensions.
In general, the divergent terms such as $t^n/\epsilon^n$ for $n\ge 2$ will be
cancelled in a renormalization procedure.
We choose $Z_{\alpha}$ to cancel the divergence as
\begin{equation}
Z_{\alpha}= 1+\frac{1}{2\epsilon}\frac{t}{2\pi}
+\frac{1}{8\epsilon^2}\left(\frac{t}{2\pi}\right)^2+\cdots
=\exp\left( \frac{t}{4\pi\epsilon} \right),
\end{equation}
near two dimensions.
We have $\mu\partial\alpha_0/\partial\mu=0$, since the bare coupling 
constant $\alpha_0$ is independent 
of the energy scale $\mu$.  This leads to
\begin{equation}
\beta(\alpha)\equiv \mu\frac{\partial\alpha}{\partial\mu}= -2\alpha
-\alpha\mu\frac{\partial \ln Z_{\alpha}}{\partial\mu}.
\end{equation}
Similarly we have
\begin{equation}
\beta(t)\equiv \mu\frac{\partial t}{\partial\mu}= (d-2)t
-t\mu\frac{\partial\ln Z_t}{\partial\mu}.
\end{equation}
Because $Z_t=1$ up to the lowest order of $\alpha$, we obtain up to the
first order of $t$, 
\begin{equation}
\mu\frac{\partial\ln Z_{\alpha}}{\partial\mu}=\frac{1}{4\pi\epsilon}
\mu\frac{\partial t}{\partial\mu}
= \mu\frac{1}{Z_{\alpha}}\frac{\partial Z_{\alpha}}{\partial\mu}
=\frac{t}{4\pi}.
\end{equation}
This results in
\begin{equation}
\beta(\alpha)= -2\alpha-\frac{1}{4\pi}\alpha t
= -2\alpha \left(1+\frac{1}{8\pi}t\right).
\end{equation}
This expression holds near two dimensions.
It then appears that $\beta(\alpha)$ has zero only at $\alpha=0$
since $t>0$, which is shown in Fig. 2.  $\beta(\alpha)$ is always negative
indicating that the asymptotic freedom is a feature of sinh-Gordon model.

\begin{figure}[htbp]
\begin{center}
\includegraphics[height=1.8cm]{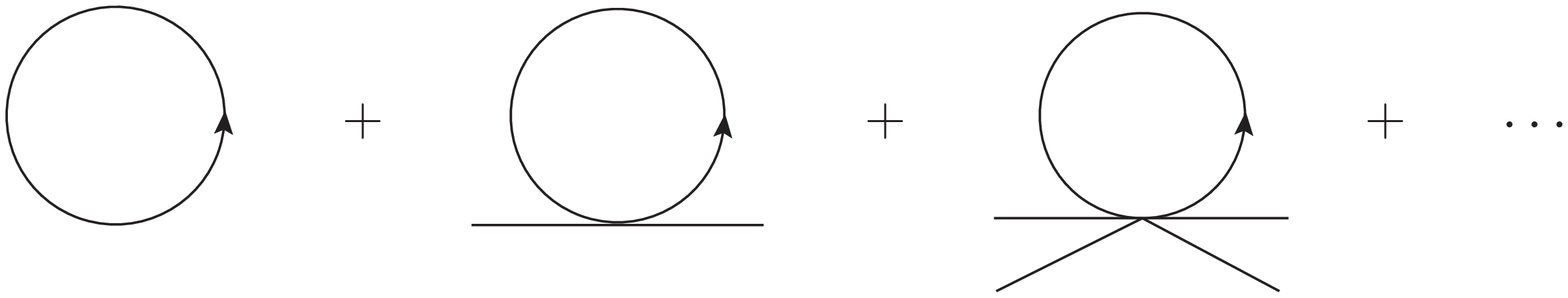}
\caption{
Tadpole diagrams for the renormalization of $\alpha$.
}
\end{center}
\end{figure}

\begin{figure}[htbp]
\begin{center}
\includegraphics[height=5.0cm]{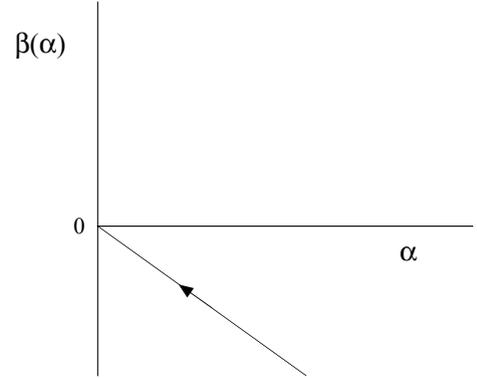}
\caption{
The beta function $\beta(\alpha)$ as a function of $\alpha$.
The arrow indicates the flow as $\mu\rightarrow \infty$.
}
\end{center}
\end{figure}

\subsection{Renormalization of $t$}

Let us examine the renormalization effect on the coupling constant $t$.
We consider the correction to the kinetic part of the action.
The correction to the action in the second order of $\alpha$ is given by
\begin{eqnarray}
S^{(2)}&=& -\frac{1}{2}\left(\frac{\mu^d}{tZ_t}\right)^2
\left(\alpha Z_{\alpha}\right)^2\int d^dx d^dx' 
 \cosh\left(\sqrt{Z_{\phi}}\phi(x)\right) \nonumber\\
&& \times \cosh\left(\sqrt{Z_{\phi}}\phi(x')\right)
\nonumber\\
&=& -\frac{1}{4}\left(\frac{\mu^d}{tZ_t}\right)^2
\left(\alpha Z_{\alpha}\right)^2\int d^dx d^dx'\Big[ \nonumber\\
&&\cosh\left(\sqrt{Z_{\phi}}(\phi(x)+\phi(x'))\right) \nonumber\\
&& +\cosh\left(\sqrt{Z_{\phi}}(\phi(x)-\phi(x'))\right)\Big].
\end{eqnarray}
The term with $\cosh\left(\sqrt{Z_{\phi}}(\phi(x)+\phi(x'))\right)$ gives an effective
potential with high-frequency mode, which is not examined in this section.
The second term $\cosh\left(\sqrt{Z_{\phi}}(\phi(x)-\phi(x'))\right)$ will give a
correction to the kinetic term, thus to the renormalization of $t$.

Based on the tadpole approximation where we consider diagrams shown in Fig. 1, 
$\cosh\left(\sqrt{Z_{\phi}}(\phi(x)-\phi(x'))\right)$ is renormalized as
\begin{eqnarray}
\cosh\left(\sqrt{Z_{\phi}}(\phi(x)-\phi(x'))\right) &\rightarrow&
\exp\left( \frac{1}{2}Z_{\phi}\langle (\phi(x)-\phi(x'))^2\rangle\right)
\nonumber\\
&\times& \cosh\left(\sqrt{Z_{\phi}}(\phi(x)-\phi(x'))\right).
\nonumber\\
\end{eqnarray}
Then we have
\begin{eqnarray}
S^{(2)}&\simeq& -\frac{1}{4}\left(\frac{\mu^d}{tZ_t}\right)^2
\left(\alpha Z_{\alpha}\right)^2\int d^dx d^dx'\Big[ \nonumber\\
&\times& \exp\left( \frac{1}{2}Z_{\phi}\langle \phi(x)^2+\phi(x')^2\rangle
-Z_{\phi}\langle\phi(x)\phi(x')\rangle\right)
\nonumber\\
&\times& \cosh\left( \sqrt{Z_{\phi}}(\phi(x)-\phi(x')) \right)\Big].
\end{eqnarray}
The correlation function $\langle\phi(x)\phi(x')\rangle$ is written as
\begin{equation}
\langle\phi(x)\phi(x')\rangle= \frac{t\mu^{2-d}Z_t}{Z_{\phi}}
\frac{\Omega_d}{(2\pi)^d}K_0(m_0|x-x'|),
\end{equation}
where $K_0(z)$ is the zeroth modified Bessel function.
$K_0(z)$ divergently increases as $z$ approaches zero.  Then,
$\exp(-Z_{\phi}\langle\phi(x)\phi(x')\rangle)$ becomes very small
when $|x-x'|$ is small.  A dominant contribution comes from the region
where $|x-x'|$ is large.  This indicates that $S^{(2)}$ gives no
contribution to the renormalization of the kinetic term since we cannot expand
$S^{(2)}$ with respect to ${\bf r}$ where ${\bf r}=x'-x$. 
Thus, there is no renormalization of $t$ up to the second order of
$\alpha$.
This is in contrast to the result for the sine-Gordon model where
$\exp(-Z_{\phi}\langle\phi(x)\phi(x')\rangle)$ becomes very large
for $x-x'\sim 0$.
The beta function for $t$ is now given as
\begin{equation}
\beta(t)= (d-2)t.
\end{equation}

\section{Wilson renormalization group method}

Let us examine the renormalization group theory for the
sinh-Gordon model based on the Wilson renormalization group method.
In Wilson's method, we start from the action
\begin{equation}
S= \int d^2x\Big[ \frac{1}{2}(\partial_{\mu}\phi(x))^2
+g\cosh(\beta\phi(x))\Big],
\end{equation}
with the cutoff $\Lambda$ in the momentum space.
We consider corrections to the action when we reduce the cutoff
from $\Lambda$ to $\Lambda-d\Lambda$.
The field $\phi(x)$ is divided into two terms $\phi(x)=\phi_1(x)+\phi_2(x)$ where
\begin{eqnarray}
\phi_1(x) &=& \int_{0\le |{\bf k}|\le \Lambda-d\Lambda}
\frac{d^2k}{(2\pi)^2}e^{i{\bf k}\cdot{\bf x}}\phi({\bf k}),\nonumber\\
\phi_2(x) &=& \int_{\Lambda-d\Lambda \le |{\bf k}|\le \Lambda}
\frac{d^2k}{(2\pi)^2}e^{i{\bf k}\cdot{\bf x}}\phi({\bf k}).
\end{eqnarray}
The action is written as
\begin{eqnarray}
S&=& \int d^2x\left[ \frac{1}{2}(\partial_{\mu}\phi_1)^2
+\frac{1}{2}(\partial_{\mu}\phi_2)^2+g\cosh(\beta(\phi_1+\phi_2))\right]
\nonumber\\
&=& S_0(\phi_1)+S_0(\phi_2)+S_1(\phi_1,\phi_2).
\end{eqnarray}
The last term $S_1$ is regarded as a perturbation.
Since the potential $\cosh\beta(\phi_1+\phi_2)$ is approximated as
\begin{eqnarray}
\cosh\beta(\phi_1+\phi_2)&=& \cosh(\beta\phi_1)\left( 1+\frac{1}{2}\beta^2\phi_2^2
+\cdots \right) \nonumber\\
&& +\sinh\beta(\phi_1)(\beta\phi_2+\cdots ),
\end{eqnarray}
the correction to the action $S_0(\phi_1)$ in the lowest order is given by
\begin{eqnarray}
S^{(1)}&=& \langle S_1\rangle
= g\int d^2 x\langle \cosh(\beta(\phi_1+\phi_2))\rangle
\nonumber\\
&\simeq& g\int d^2 x
\exp\left( \frac{\beta^2}{2}\langle\phi_2^2\rangle\right)\cosh(\beta\phi_1)
\nonumber\\
&\simeq& g\left( 1+\frac{\beta^2}{4\pi}
\frac{d\Lambda}{\Lambda}\right)\int d^2x\cosh(\beta\phi_1),
\end{eqnarray}
where $\langle\cdots\rangle$ indicates the expectation value with
respect to the action $S_0(\phi_2)$:
\begin{equation}
\langle Q\rangle= \frac{1}{Z_2}\int d\phi_2 Qe^{-S_0(\phi_2)},
\end{equation}
with $Z_2=\int d\phi_2 e^{-S_0(\phi_2)}$.
$\langle\phi_2^2\rangle$ reads $\langle\phi_2(x)\phi_2(x')\rangle$
for $x'=x$ where the correlation function in this formulation is
\begin{eqnarray}
\langle\phi_2(x)\phi_2(x')\rangle &=& 
\int_{\Lambda-d\Lambda\le |{\bf k}|\le\Lambda} \frac{d^2k}{(2\pi)^2}
e^{i{\bf k}\cdot (x-x')}\langle\phi_2({\bf k})\phi_2(-{\bf k})\rangle
\nonumber\\
&=& \frac{1}{2\pi}J_0(\Lambda |x-x'|)\frac{d\Lambda}{\Lambda},
\end{eqnarray}
where $\langle\phi_2({\bf k})\phi_2(-{\bf k})\rangle=1/k^2$ and $J_0(z)$
is the zeroth Bessel function.

The correction to $S_0(\phi_1)$ in the second order of $\alpha$ is given by
\begin{eqnarray}
S^{(2)}&=& -\frac{1}{2}g^2\int d^2x d^2x'
\Big[ \nonumber\\
&&  \langle\cosh(\beta(\phi_1(x)+\phi_2(x)))\cosh(\beta(\phi_1(x')+\phi_2(x')))\rangle
\nonumber\\
&& -\langle\cosh(\beta(\phi_1(x)+\phi_2(x)))\rangle
\langle\cosh(\beta(\phi_1(x')+\phi_2(x')))\rangle \Big]
\nonumber\\
&\simeq& -\frac{1}{4}g^2
\int d^2x d^2x' \exp\left(\beta^2\langle\phi_2(x)^2\rangle\right) \Big[ \nonumber\\
&& \big[\exp\left(\beta^2\langle\phi_2(x)\phi_2(x')\rangle\right)-1\big]
\cosh(\beta(\phi_1(x)+\phi_1(x'))) \nonumber\\
&& + \big[ \exp\left(-\beta^2\langle\phi_2(x)\phi_2(x')\rangle\right)-1\big]
\nonumber\\
&& \times \cosh(\beta(\phi_1(x)-\phi_1(x'))) \Big].
\end{eqnarray}
The first term with $\cosh(\beta(\phi_1(x)+\phi_1(x')))$ gives a potential of the form
$\cosh(2\phi_1)$ which we neglect in this section.
$\big[ \exp\left(-\beta^2\langle\phi_2(x)\phi_2(x')\rangle\right)-1\big]$ in the 
second term becomes
small as $|x-x'|$ decreases.
Hence there is also no renormalization effect for the coupling constant $t$ 
up to the order of $\alpha^2$ in this formulation.
Then, the effective action for the field $\phi_1$ with the cutoff $\Lambda-d\Lambda$ is
\begin{eqnarray}
S_{\Lambda-d\Lambda}&=& \int d^2x \left[ \frac{1}{2}(\partial_{\mu}\phi_1)^2
+g\left( 1+\frac{\beta^2}{4\pi}\frac{d\Lambda}{\Lambda}\right)
 \cosh(\beta\phi_1) \right].
\nonumber\\
\end{eqnarray}
We perform the scale transformation to let the cutoff be $\Lambda$:
\begin{eqnarray}
x'&=& e^{-d\ell}x, \\
{\bf k}' &=& e^{d\ell}{\bf k} ,\\
\phi_1({\bf k}) &=& \zeta \tilde{\phi}_1({\bf k}),
\end{eqnarray}
where $d\ell=d\Lambda/\Lambda$.  We have
\begin{eqnarray}
\phi_1(x) &=& \zeta e^{-2d\ell}\int_{0\le |{\bf k}'|\le\Lambda}
\frac{d^2k'}{(2\pi)^2}e^{i{\bf k}'\cdot x'}\tilde{\phi}_1({\bf k}')
\nonumber\\
&=& \zeta e^{-2d\ell}\tilde{\phi}_1(x').
\end{eqnarray}
The effective action reads
\begin{eqnarray}
S_{\Lambda-d\Lambda}&=& \int d^2 x'\Big[ \zeta^2 e^{-4d\ell}\frac{1}{2}
(\partial_{\mu}\tilde{\phi}_1(x'))^2  \nonumber\\ 
&& +e^{2d\ell}g\left( 1+\beta^2\frac{1}{4\pi}
\frac{d\Lambda}{\Lambda}\right) \cosh(\beta\zeta e^{-2d\ell}\tilde{\phi}_1(x'))\Big]
\nonumber\\
&=& \int d^2x'\Big[ \frac{1}{2}(\partial_{\mu}\tilde{\phi}_1(x'))^2 
 +g\left( 1+2\frac{d\Lambda}{\Lambda}
+\beta^2\frac{1}{4\pi}\frac{d\Lambda}{\Lambda}\right) \nonumber\\
&& \times \cosh(\beta\tilde{\phi}_1(x')\Big],
\end{eqnarray}
where we set $\zeta=e^{2d\ell}$.
This indicates that $t$ is not renormalized and $g$ is renormalized
to $g+dg$:
\begin{equation}
\Lambda\frac{dg}{d\Lambda} = 2g+\frac{1}{4\pi}\beta^2 g.
\end{equation} 
The equation for $\alpha$ reads 
\begin{equation}
\Lambda\frac{d\alpha}{d\Lambda} = 2\alpha+\frac{1}{4\pi}t\alpha.
\end{equation} 
Hence, we obtained the same equation as in the previous section and 
the effective $\alpha$ increases as the cutoff $\Lambda$ decreases
to the low-energy region.
Because we examined the derivative in the descent direction for
$\Lambda\rightarrow \Lambda-d\Lambda$, the above equation has opposite
sign to the equation in Eq. (18).  Since $d\Lambda$ is related to $d\mu$ as
$d\ln\Lambda=-d\ln\mu$, we have the same equation.
Since the beta functions in the dimensional regularization method
were obtained by fixing bare quantities, we used a partial derivative
expression in Sect. 3.

\section{Renormalization group flow}

The renormalization group equations in the lowest order of $\alpha$ and $t$ are
\begin{eqnarray}
\mu\frac{\partial\alpha}{\partial\mu}&=& -2\alpha\left(1+\frac{1}{8\pi}t\right),
\nonumber\\
\mu\frac{\partial t}{\partial\mu}&=& (d-2)t.
\end{eqnarray}
In two dimensions ($d=2$), we have $t=t_0=$ constant and
\begin{equation}
\alpha= \alpha_0\left(\frac{\mu}{\mu_0}\right)^{-(2+t_0/4\pi)},
\end{equation}
where we adopt that $\alpha=\alpha_0$ at $\mu=\mu_0$.
The flow as $\mu\rightarrow\infty$ is shown in Fig. 3.
$t$ remains constant, and $\alpha$ decreases and approaches zero.

For $d$ near 2, the solution is given by
\begin{equation}
t= t_0\left(\frac{\mu}{\mu_0}\right)^{d-2},
\end{equation}
\begin{equation}
\alpha= \alpha_0\left(\frac{\mu}{\mu_0}\right)^{-2}\exp\left[
-\frac{t_0}{4\pi}\frac{1}{d-2}\left( \left(\frac{\mu}{\mu_0}\right)^{d-2}-1\right)
\right],
\end{equation}
where $t_0$ and $\alpha_0$ are initial values of $t$ and $\alpha$.
In the limit $d\rightarrow 2$, this set of solutions reduces to that
for $d=2$.
The renormalization group flow is shown in Fig. 4 for general $d$.
In the high-energy region the effective $\alpha$ becomes vanishingly small,
and instead in the low-energy region $\alpha$ becomes very large and will dominate
the low-energy property.
The function in Eq. (43) is important in the theory of the sinh-Gordon
model.  In a variational theory\cite{ing86}, the expectation value of
the potential energy is expressed with this function.

\begin{figure}[htbp]
\begin{center}
\includegraphics[height=6.0cm]{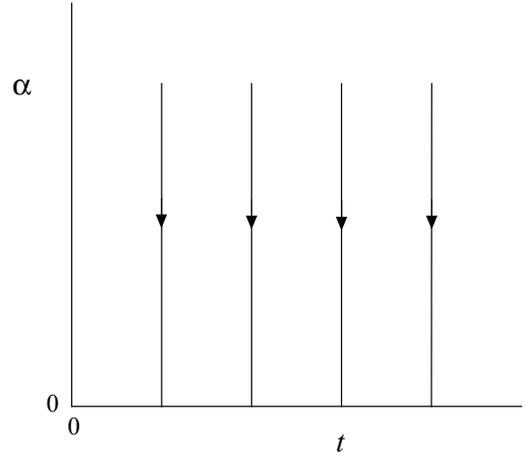}
\caption{
The renormalization group flow  in two dimensions $d=2$ where
the arrow indicates the flow as $\mu\rightarrow \infty$.
}
\end{center}
\end{figure}

\begin{figure}[htbp]
\begin{center}
\includegraphics[height=7.5cm]{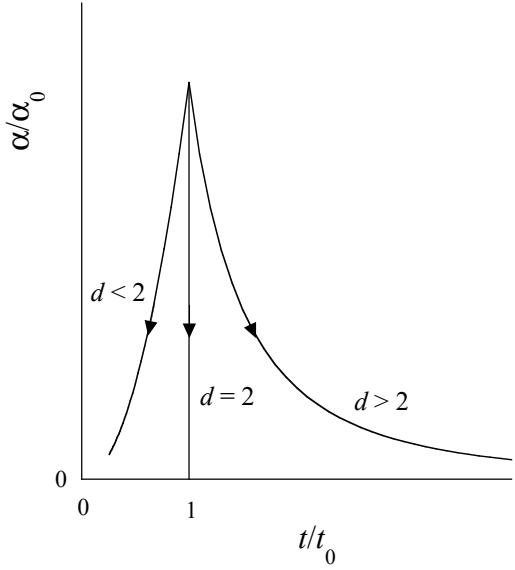}
\caption{
The renormalization group flow starting from $t=t_0$ and
$\alpha=\alpha_0$ for $d>2$, $d=2$ and $d<2$, respectively, where
the arrow indicates the flow as $\mu\rightarrow \infty$.
}
\end{center}
\end{figure}

\section{Renormalization of high-frequency modes}

In the renormalization procedure in the previous section
a high-frequency mode such as $\cosh(2\phi)$ appears that we have neglected so far.
We examine the renormalization of high-frequency terms in this section.
Let us consider the action:
\begin{eqnarray}
S&=& \int d^2 x \left[ \frac{1}{2}(\partial_{\mu}\phi)^2
+\sum_{n=1} g_n\cosh(n\beta\phi)\right]\nonumber\\
&=& \int d^2x \Big[ \frac{1}{2}(\partial_{\mu}\phi_1)^2
+\frac{1}{2}(\partial_{\mu}\phi_2)^2 \nonumber\\
&& +\sum_{n=1}g_n\cosh(n\beta (\phi_1+\phi_2))\Big]\nonumber\\
&=& S_0(\phi_1)+S_0(\phi_2)+S_1(\phi_1,\phi_2),
\end{eqnarray}
where $\phi_1$ and $\phi_2$ were defined in Sect. 4. 
$g_n\equiv\alpha_n/t$ ($n=1,2,...$) are the coupling constants
and $\beta=\sqrt{t}$.
The lowest-order correction in the Wilson method is
\begin{eqnarray}
S^{(1)}&=& \sum_n g_n\exp\left(n^2\frac{\beta^2}{2}\langle\phi_2^2\rangle\right)
\int d^2 x\cosh(n\beta\phi_1(x))\nonumber\\
&=& \sum_n g_n\exp\left(n^2\frac{\beta^2}{4\pi}\frac{d\Lambda}{\Lambda}\right)
\int d^2 x\cosh(n\beta\phi_1(x)).
\nonumber\\
\end{eqnarray}
The second-order correction is given as
\begin{eqnarray}
S^{(2)}&=& -\frac{1}{4}\sum_{nm}g_ng_m
\exp\left((n^2+m^2)\frac{\beta^2}{4\pi}\frac{d\Lambda}{\Lambda}\right)
\nonumber\\
&& \times\int d^2x d^2x'
\left( \exp\left(nm\beta^2\langle\phi_2(x)\phi_2(x')\rangle\right)-1 \right)
\nonumber\\
&& \times \cosh(n\beta\phi_1(x)+m\beta\phi_1(x'))\nonumber\\
&\simeq& -\frac{1}{4}\sum_{nm}nmg_n g_m\int d^2 r
\frac{\beta^2}{2\pi}J_0(\Lambda r)\frac{d\Lambda}{\Lambda} \nonumber\\
&& \times \int d^2 x\cosh((n+m)\beta\phi_1(x))\nonumber\\
&=& -\frac{C_{\Lambda}}{4}\beta^2\sum_{nm}nmg_ng_m
\frac{d\Lambda}{\Lambda}\int d^2x\cosh((n+m)\beta\phi_1(x)),
\nonumber\\
\end{eqnarray}
where $C_{\Lambda}$ is a constant:
\begin{equation}
C_{\Lambda}= \int d^2 r\frac{1}{2\pi}J_0(\Lambda r).
\end{equation}
Then, the effective action is
\begin{eqnarray}
S_{\Lambda-d\Lambda}&=& \int d^2 x \Big[ \frac{1}{2}(\partial_{\mu}\phi_1)^2
+\sum_{n=1}g_n\left( 1+\frac{n^2\beta^2}{4\pi}\frac{d\Lambda}{\Lambda}\right)
\cosh(n\beta\phi_1)
\nonumber\\
&-& \frac{C_{\Lambda}}{4}\beta^2\sum_{nm,1\le m\le n-1}g_{n-m}g_m(n-m)m
\frac{d\Lambda}{\Lambda}\cosh(n\beta\phi_1) \Big].
\nonumber\\
\end{eqnarray}
After the scaling transformation $x\rightarrow x'=e^{-d\ell}x$, this action
results in the scaling equations
\begin{eqnarray}
\Lambda\frac{dg_n}{d\Lambda}&=&  \left( 2+n^2\frac{\beta^2}{4\pi}\right)g_n
-\frac{C_{\Lambda}}{4}\beta^2\sum_{1\le m\le n-1}m(n-m)g_{n-m}g_m,\nonumber\\
\end{eqnarray}
\begin{equation}
\Lambda\frac{d\beta}{d\Lambda} = 0.
\end{equation}
Since $g_n=\alpha_n/t$, we have
\begin{equation}
\Lambda\frac{d\alpha_n}{d\Lambda}=  \left( 2+n^2\frac{t}{4\pi}\right)\alpha_n
-\frac{C_{\Lambda}}{4}\sum_{1\le m\le n-1}m(n-m)\alpha_{n-m}\alpha_m.
\end{equation}
For small $n$  ($n=1,2,...$), the equations for $\alpha_n$ are
\begin{eqnarray}
\Lambda\frac{d\alpha_1}{d\Lambda} &=& \left(2+\frac{t}{4\pi}\right)\alpha_1, \\
\Lambda\frac{d\alpha_2}{d\Lambda} &=& \left(2+\frac{t}{\pi}\right)\alpha_2
-\frac{C_{\Lambda}}{4}\alpha_1^2,\\
\Lambda\frac{d\alpha_3}{d\Lambda} &=& \left(2+\frac{9t}{4\pi}\right)\alpha_3
-C_{\Lambda}\alpha_1\alpha_2.
\end{eqnarray}
$\alpha_1$ obviously decreases to zero in the high-energy region.
Thus, $\alpha_2$ and $\alpha_3$ also decreases as the energy scale increase.
Hence the sinh-Gordon theory with high-frequency modes remains an
asymptotically free theory.

\section{Summary}

We have presented a renormalization group analysis of the sinh-Gordon model.
The analysis is based on the dimensional regularization method and also
the Wilson renormalization group method.
A set of beta functions were derived and its scaling property was discussed. 
In contrast to the sine-Gordon model, the sinh-Gordon model exhibits an
asymptotic freedom with vanishing $\alpha$ in the limit $\mu\rightarrow \infty$ 
in two dimensions ($d=2$).
Up to the second order of $\alpha$, the coupling constant $t$ is not
renormalized in two dimensions.
In Ref.\cite{ing86}, the ground-state energy was estimated by using
a variational wave function.  The same function as in Eq. (43)
appears in the evaluation of the potential energy and plays an important
role.  This indicates
that the two results are consistent. 
We have also examined the generalized model with interactions
$\alpha_n\cosh(n\beta\phi)$ ($n=1,2,...$).  It was shown
based on the Wilson method that the
equation for $\alpha_1$ remains the same and that the generalized
sinh-Gordon theory is an asymptotically free theory.
We obtain the same renormalization group equations by employing the
dimensional regularization method, where the coefficients for higher-order
terms are slightly modified.

The sinh-Gordon model belongs to a universality class showing asymptotic
freedom.  The nonlinear sigma model and non-Abelian Yang-Mills theory
also belong in this class.
Thus the sinh-Gordon model is an interesting model and may be applied
to various phenomena in the future.
In the infrared region, the parameter $\alpha$ increases and dominates
the property of the system.  The physical property is determined by the
potential energy $\cosh\phi$.  In this region, $\phi$ may be small since
the kinetic term is negligibly small.  Then, $\cosh\phi$ can be expanded
by $\phi$ to investigate the low-energy property.  The effective action is
given by a $\phi^4$ theory:
\begin{equation}
S_{low}= \int d^2 x \left[ \frac{1}{2}(\partial_{\mu}\phi)^2
+\frac{\alpha}{2}\phi^2+\frac{1}{4!}t\alpha\phi^4 \right],
\end{equation}
where we did the scale transformation $\phi\rightarrow \sqrt{t}\phi$.

\section{Acknowledgment}
This work was supported by Grant-in-Aid from the Ministry of Education,
Culture, Sports, Science and Technology (MEXT) of Japan (no. 17K05559).

\end{document}